\documentclass{PoS}
\usepackage{graphicx}

\usepackage{color}

\newcommand{\mST}{m_{S,T}{}}
\newcommand{\mSV}{m^{0}_{S}{}}
\newcommand{\ysx}{y_{\chi}}

\newcommand{\Ycr}{ Y_{ \rm DM, 0 }{} }
\newcommand{\Yc}{ Y_{\rm DM}{} }
\newcommand{\mc}{ m_{\chi} }

\newcommand{\zR}{ z_{\rm RH} }
\newcommand{\xR}{ x_{\rm RH} }

\newcommand{\ie}{{\em i.e.}  }
\newcommand{\eg}{{\em e.g.}  }

\newcommand{\GeV}{{\rm GeV}}

\newcommand{\vev}[1]{\langle #1 \rangle}
\newcommand{\lrb}[1]{\left( #1 \right)}

\usepackage{ifthen}
\usepackage{tikz}

\newcommand{\eqs}[1]{\foreach\i[count=\NumArgs] in {#1}{}%
	\ifthenelse{\equal{\NumArgs}{1}}{eq.~(\ref{#1})}%
	{\ifthenelse{\equal{\NumArgs}{2}}%
		{eqs.~\foreach\i[count=\q]in{#1}{\ifthenelse{\equal{\q}{\NumArgs}}{and (\ref{\i})}{(\ref{\i})~}}}%
		{eqs.~\foreach\i[count=\q]in{#1}{\ifthenelse{\equal{\q}{\NumArgs}}{and (\ref{\i})}{(\ref{\i}),~}}}}}

\newcommand{\Eqs}[1]{\foreach\i[count=\NumArgs] in {#1}{}%
	\ifthenelse{\equal{\NumArgs}{1}}{eq.~(\ref{#1})}%
	{\ifthenelse{\equal{\NumArgs}{2}}%
		{Eqs.~\foreach\i[count=\q]in{#1}{\ifthenelse{\equal{\q}{\NumArgs}}{and (\ref{\i})}{(\ref{\i})~}}}%
		{Eqs.~\foreach\i[count=\q]in{#1}{\ifthenelse{\equal{\q}{\NumArgs}}{and (\ref{\i})}{(\ref{\i}),~}}}}}

\newcommand{\refs}[1]{\foreach\i[count=\NumArgs] in {#1}{}%
	\ifthenelse{\equal{\NumArgs}{1}}{(\ref{#1})}%
	{\ifthenelse{\equal{\NumArgs}{2}}%
		{\foreach\i[count=\q]in{#1}{\ifthenelse{\equal{\q}{\NumArgs}}{and (\ref{\i})}{(\ref{\i})~}}}%
		{\foreach\i[count=\q]in{#1}{\ifthenelse{\equal{\q}{\NumArgs}}{and (\ref{\i})}{(\ref{\i}),~}}}}}

\usepackage{amsmath}

\title{Forbidden Freeze-In}
\ShortTitle{ Forbidden Freeze-In}

\author{ L. Darm\'e\\
	 National Centre for Nuclear Research, ul. Pasteura 7, 02-093 Warsaw, Poland\\
	E-mail: \email{Luc.Darme@ncbj.gov.pl}}

\author{ A. Hryczuk\\
	National Centre for Nuclear Research, ul. Pasteura 7, 02-093 Warsaw, Poland\\
	E-mail: \email{Andrzej.Hryczuk@ncbj.gov.pl}}

\author{ \speaker{D. Karamitros}\\
	National Centre for Nuclear Research, ul. Pasteura 7, 02-093 Warsaw, Poland\\
	E-mail: \email{Dimitrios.Karamitros@ncbj.gov.pl}}

\author{ L. Roszkowski\\
	National Centre for Nuclear Research, ul. Pasteura 7, 02-093 Warsaw, Poland,\\
	Astrocent,  Nicolaus  Copernicus  Astronomical  Center  Polish  Academy  of  Sciences,  Bar-tycka 18, 00-716 Warsaw\\
	E-mail: \email{Leszek.Roszkowski@ncbj.gov.pl}}

\FullConference{Corfu Summer Institute 2019 "School and Workshops on Elementary Particle Physics and Gravity" (CORFU2019)\\
	31 August - 25 September 2019\\
	Corfù, Greece}

\abstract{We study the importance of a frozen-in dark matter production regime, where the dark matter particle is produced via kinematically forbidden decays
	 that arise from  significant thermal correction to the mass a mediator particle in the plasma.}

\begin{document}
  
\section{Introduction}

The existence of dark matter (DM) in the form of particle(s) that interact predominantly via gravitational interactions is a well established fact. 
There is a number of production mechanisms that can produce the observed DM relic abundance (depending on the model),
from which a particularly interesting one is the so-called ``freeze-in" mechanism~\cite{Ellis:1984eq,Covi:2001nw,Hall:2009bx}. 
The basic assumption for this mechanism is that the DM particle is not produced with the rest of the plasma (\eg negligible decay rate of inflaton to DM) and is unable to reach thermal equilibrium due to very weak interaction strength with the plasm.

In this proceedings, we present our work~\cite{Darme:2019wpd} on a (largely) neglected case of freeze-in DM production  where the DM particle ($\chi$) is produced via the decay
of a plasma particle ($S$) which develops a substantial thermal mass. The main point is that at high enough temperatures, the thermal correction to the  mass of $S$ becomes 
large resulting to  kinematically forbidden decay $S \to \bar{\chi} \chi$.~\footnote{This production case was first noted in~\cite{Rychkov:2007uq} for gravitino production.}
In order to study this case as generally as possible, in the following sections we focus on a simple Higgs portal model, and take a closer look at the  ``forbidden freeze-in" regime to identify its main phenomenological features.

\section{Standard Freeze-in}
In this section we summarize the case of standard freeze-in, where we note that there are two possibilities. The first one is the DM production via renormalizable operators, and
the second on is the DM production via higher dimension operators.

\subsubsection*{ Production via renormalizable operators}
The simplest possibility in this case is the DM production via decays of some mediator particle ($S$). In the standard freeze-in the thermal correction to the mass of $S$ is 
neglected, \ie we study the case where  the decay $S \to \bar{\chi} \chi$ is only allowed in the vacuum.
Assuming that this the decay channel dominates the DM production, we  estimate the DM relic abundance  from the Boltzmann equation (BE) for the yield ($Y \equiv n/s$) which takes the form
\begin{equation}
-HsT \, \delta_{h}^{-1} \frac{d\Yc}{dT}=\dfrac{  \Gamma_\chi}{ \pi^2} \, K_{1}( m_{S}/T) \; m_{S}^{2} \; T \, ,
\label{eq:BE_std}
\end{equation}
with $\Gamma_\chi$ the decay width of $S$ to a pair of DM particles. The BE can be approximately solved by~\footnote{The relativistic degrees of freedom ($g$ and $h$) are evaluated at the mean  $x$ defined by 
	$\vev{x}\equiv \dfrac{\int_{0}^{\infty} dx \, x^{3} K_{1}(x) \times x }{\int_{0}^{\infty} dx \, x^{3} K_{1}(x)}  \approx 3.4 \, .$
}
\begin{align}
\Ycr \approx& \lrb{ \dfrac{\Gamma_\chi (m_S,m_\chi) }{ 1.6 \times 10^{-36} \, \GeV} } \lrb{ \dfrac{1 \  \GeV}{m_S} }^2 
m_{\chi} \lrb{ \dfrac{1}{ \sqrt{g} \; h }  } \! \Big{|}_{x=\vev{x}} \, .
\label{eq:YIR_std}
\end{align}
We note that in this csae, the production is dominated at temperatures somewhere below the mass of $S$. This is because, the number density of $S$ becomes exponentially suppressed as $S$ becomes non-relativistic, which happens around $T \approx m_S/3$.

\subsubsection*{ Production via non-renormalizable operators}
If the DM production occurs via non-renormalizable operator (of dimension $d$), we expect the production to be dominated at high temperatures. Focusing on the simplest possibility of production via $2  \to 2$ processes, the amplitude  for such processes in the high energy (\ie high temperature) approximation is    
$$	|\mathcal{M}|^{2} \approx \gamma_d \lrb{\dfrac{\sqrt{\hat{s}}}{\Lambda}}^{2n} \;, $$  
with $n=d-4$.  Following~\cite{Hall:2009bx}, we can show that the yield becomes
\begin{align}
\Ycr \approx \dfrac{\xR^{1-2n} - x_{0}^{1-2n}  }{ 2n-1} \lrb{ \dfrac{  4^{n} n!  (n+1)!  \gamma_{d}}{2.34 \times 10^{-15}} }   
\lrb{\dfrac{m_{S}}{\Lambda}}^{2n}  \lrb{\dfrac{1 \ \GeV}{m_{S}} } \lrb{ \dfrac{1}{ \sqrt{g}  h }  } \! \Big{|}_{x \sim \xR }\, ,
\label{eq:YUV_std}
\end{align}
which shows that, as expected, the production for $d>4$ is dominated at high temperatures. We note that for $d \leq 4$, the production is dominated at low temperature, where the masses of the particles involved become important, and~\eqs{eq:YUV_std} cannot be applied.

\begin{figure}[htb]
\centerline{%
\includegraphics[width=0.5\textwidth]{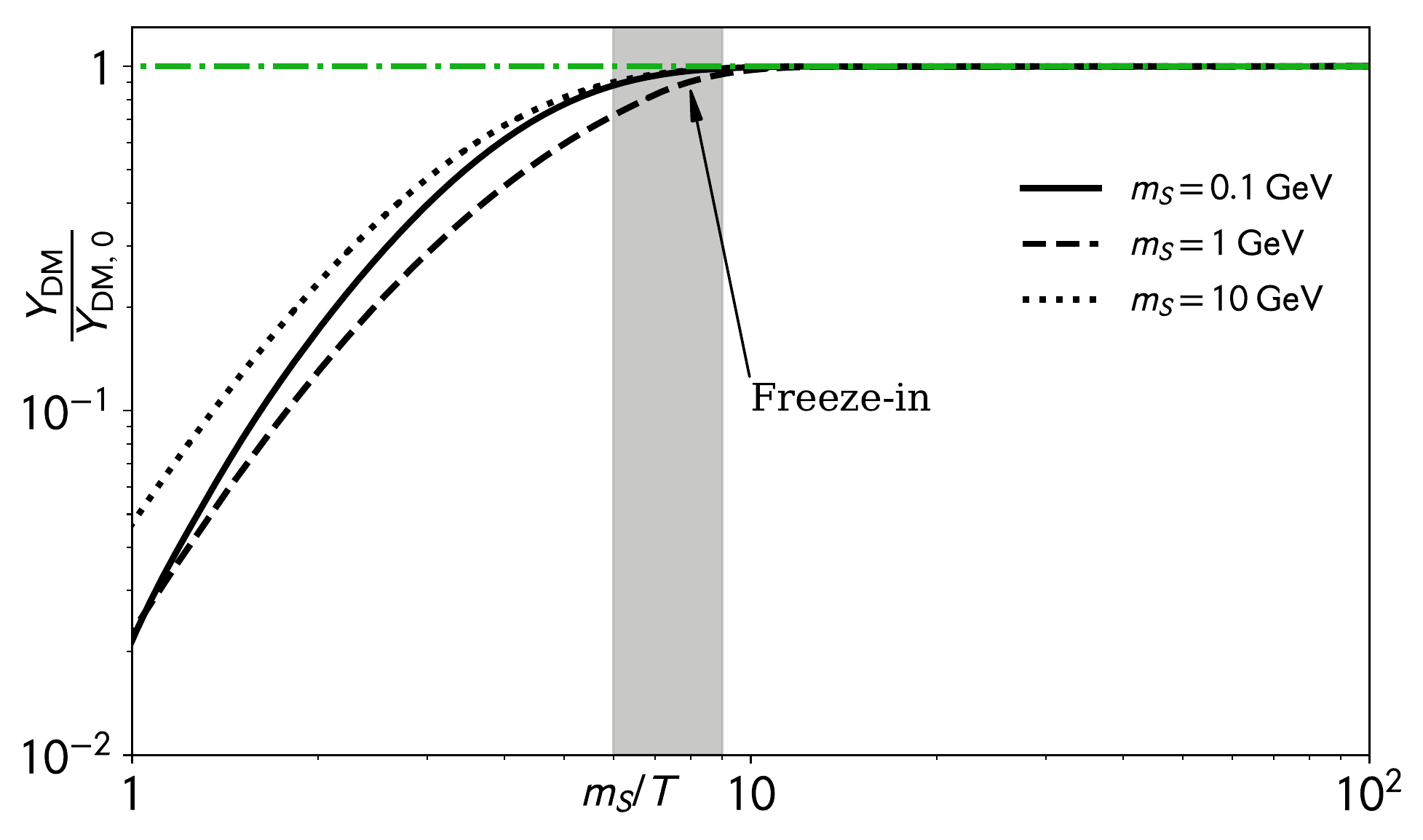}
\includegraphics[width=0.5\textwidth]{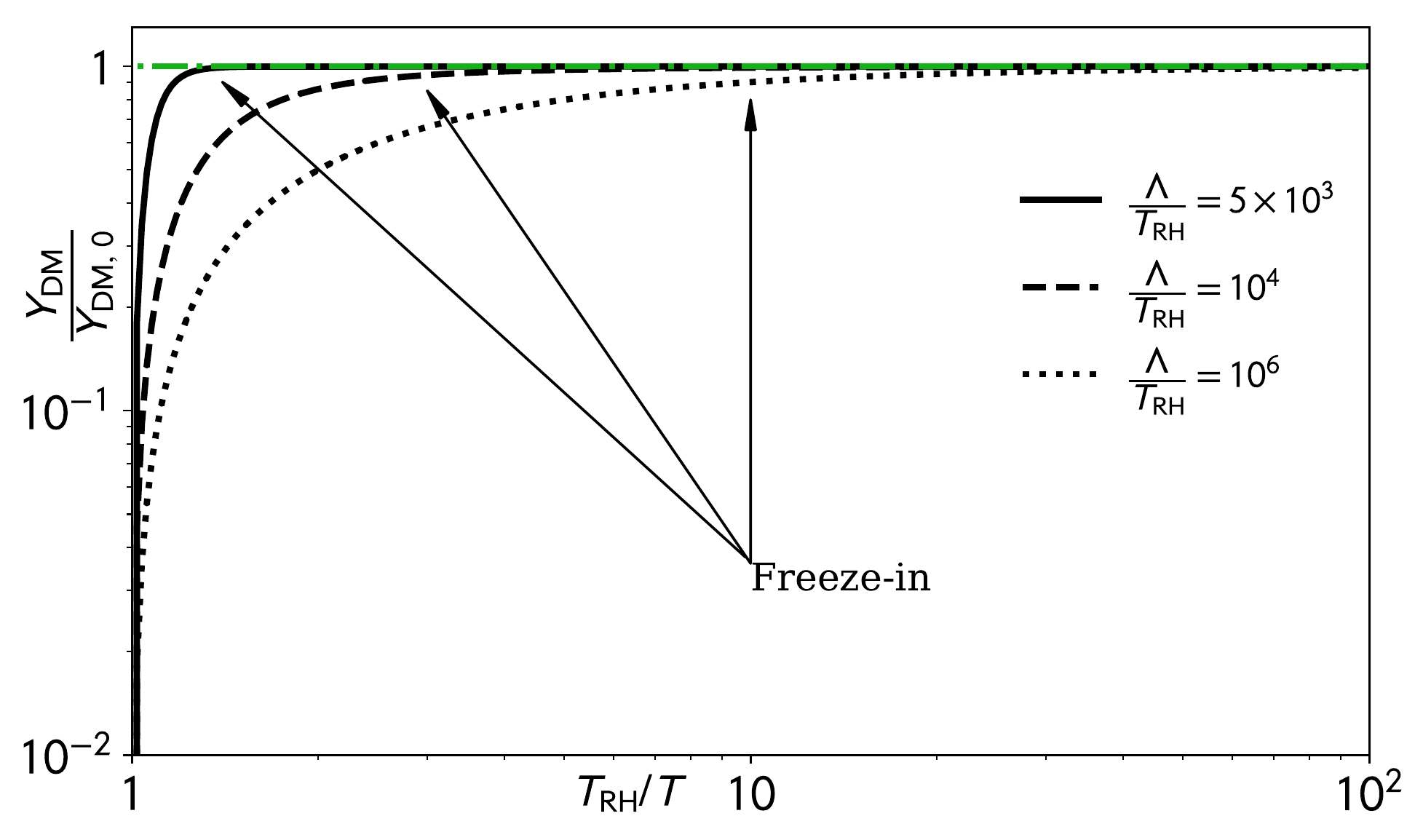}}
\caption{A typical evolution of the DM yield for the standard freeze-in via renormalizable (left) and non-renormalizable (right) operators. }
\label{Fig:Y_std}
\end{figure}
The two solutions of the relevant BEs are shown in Fig.~\ref{Fig:Y_std}. The left figure shows the evolution of the yield assuming the renormalizable  interaction term
$\mathcal{L}_{\rm int}= - \ysx S \bar{\chi}\chi$, while for the right one we assume an operator of the form $\mathcal{L}_{\rm int}= - \dfrac{1}{\Lambda} SS \bar{\chi}\chi$.
As already mentioned, we observe  that the number of DM particles (which is proportional to the yield) freezes-in for temperatures below the mass of the mediator (reheating temperature) for production due to renormalizable (non-renormalizable) operator.

\section{Forbidden Freeze-in}
This section is dedicated to the production of DM via kinematically forbidden decays, \ie forbidden freeze-in.~\footnote{We note that only decays can be kinematically forbidden, 
since for more that one initial state particles the center-of-mass energy increases with temperature.} We should note that the production via forbidden decays is a generic feature of  freeze-in, since the DM is produced by particles in thermal contact with the plasma which develop thermal corrections to their masses. So, clearly there should be a region in the parameter space that is not accessible via the standard freeze-in, and opens-up only when we consider the forbidden regime.

In order to focus on the forbidden regime, we assume that DM  is a Dirac fermion and the mediator is a scalar singlet particle ($S$) that interacts with the 
Standard Model (SM) via the Higgs.~\footnote{Since the forbidden freeze-in is a generic feature of the freeze-in mechanism, we expect the qualitative results to hold for other models, too.} That is, the relevant Lagrangian terms are
\begin{equation}
\mathcal{L}\supset - \ysx    \ S \, \bar{\chi} \chi \;  -\dfrac{\lambda_S}{4!} S^{4} - \dfrac{1}{2} \mSV^{2} S^{2} -\mc \bar{\chi} \chi \;+ \mathrm{( SH-terms)}\, .
\label{eq:LSchi}
\end{equation}
For the moment, since we are interested in showing how  forbidden freeze-in works, we assume that $S$ is in equilibrium via some interaction with the Higgs. We will specify the $S$ interactions with the Higgs later, where we study its phenomenology. 
Without great loss of generality we assume that the total mass of $S$ is given by
\begin{equation}
\mST^2 \approx \mSV^2 + \alpha^2 T^{2} \;,
\label{eq:mST}
\end{equation}
where assuming that the self-interaction dominates the thermal mass correction, $\alpha^2 = \dfrac{\lambda_S}{24}$.~\footnote{ If there are other interactions that affect the thermal mass, $\alpha$ gets simply shifted.}

\subsubsection*{Forbidden freeze-in via renormalizable operators}
The BE for the production of DM is given by~\eqs{eq:BE_std} with $m_S \to \mST$. However, if $\mc \gg \mSV$, the production happens only at high-temperature and the BE can be simplified to  
\begin{eqnarray}
\dfrac{d\Yc}{dz} \approx \lrb{ \dfrac{\Gamma_\chi  }{ 5.93 \times 10^{-19} \, \GeV} }  \lrb{ \dfrac{1 \ \GeV}{2 \mc} }^2 \dfrac{\alpha^4 \; K_{1}(\alpha)}{ \sqrt{g}\; h } \delta_{h} \; z \;,
\label{eq:BE_FFI}
\end{eqnarray}
where $z \equiv \dfrac{2\mc}{\alpha  T}$ and $$\Gamma_\chi \approx \frac{\ysx^{2}}{4 \pi} \dfrac{\lrb{  1 - z^2 }^{3/2}}{z} \mc  \,.$$ 
Since at  $\mST = 2 \mc$ the production stops, we integrate this BE from $\zR \to \infty$ to $z=1$, which gives the solution~\footnote{The mean $z$ is defined as
	$
	\vev{z}\equiv \dfrac{\int_{0}^{1} dz \; \lrb{ 1-z^{2} }^{3/2} \times z }{\int_{0}^{1} dz \,\lrb{ 1-z^{2} }^{3/2} }  \approx 0.34 \, .
	$
}
\begin{equation}
	\Ycr= \lrb{\dfrac{\alpha^2 \; \ysx }{ 5 \times 10^{-9}}}^{2} \lrb{ \dfrac{1 ~\GeV}{2\mc} } \; K_{1}(\alpha) \;  \lrb{ \dfrac{  1 }{ \sqrt{g}\; h } }_{z=\vev{z}}  \;.
\label{eq:YIR_FFI}
\end{equation} 
We should point out that in this case the freeze-in temperature is dictated by the DM mass, while in the standard treatment it is typically around the mass of the mediator.~\footnote{The freeze-in temperature can depend on the DM mass in the standard treatment only if the DM is produced by some other process. For example, if the production channel is $SS \to \chi \chi$ and $\mc \gg m_S$, the production should stop around $T \sim \mc$, where the energy carried by $S$ is not enough to produce more DM particles. }
We also note that the yield today is proportional to $\alpha^4 \ysx^2$. This makes the forbidden freeze-in very inefficient for typical couplings for frozen-in DM. So we expect 
the forbidden freeze-in regime to open-up new regions in the coupling $\ysx$ along with the new mass ranges. These are two key features of the forbidden freeze-in scenario.

\subsubsection*{Forbidden freeze-in via non-renormalizable operators}
In analogy to the standard freeze-in case, we may define a decay width due to some non-renormalizable operator as
$$\Gamma_\chi \sim  \dfrac{\gamma_{S \chi}}{16 \pi} \;\alpha^{2n+1}  \;
\lrb{\frac{T}{\Lambda}}^{2n} T  \, .$$
Then, the BE of~\eqs{eq:BE_FFI} still applies, but the solution becomes
\begin{equation}
	\Ycr= 
\frac{\zR^{1-2n} -1 }{2n-1}   \lrb{ \dfrac{\alpha^4 \; K_{1}(\alpha) \; \gamma_{S \chi}  }{ 2.96 \times 10^{-17} } }  \lrb{ \dfrac{2 \mc}{\Lambda} }^{2n}  
\lrb{\dfrac{1 \ \GeV}{2 \mc} } \lrb{ \dfrac{1}{ \sqrt{g} \; h }  } \! \Big{|}_{z \sim \zR }\;,
\label{eq:YUV_FFI}
\end{equation} 
where again we find that the forbidden freeze-in via non-renormalizable operators is dominated around the reheating temperature. Also, as in the standard case, for $d \leq 4$ we observe that the production is dominated at low temperatures (close to $z=1$), where the high temperature approximation for $\Gamma_\chi$ breaks down and the solution is given by~\eqs{eq:YIR_FFI}.

\begin{figure}[htb]
	\centerline{%
		\includegraphics[width=0.5\textwidth]{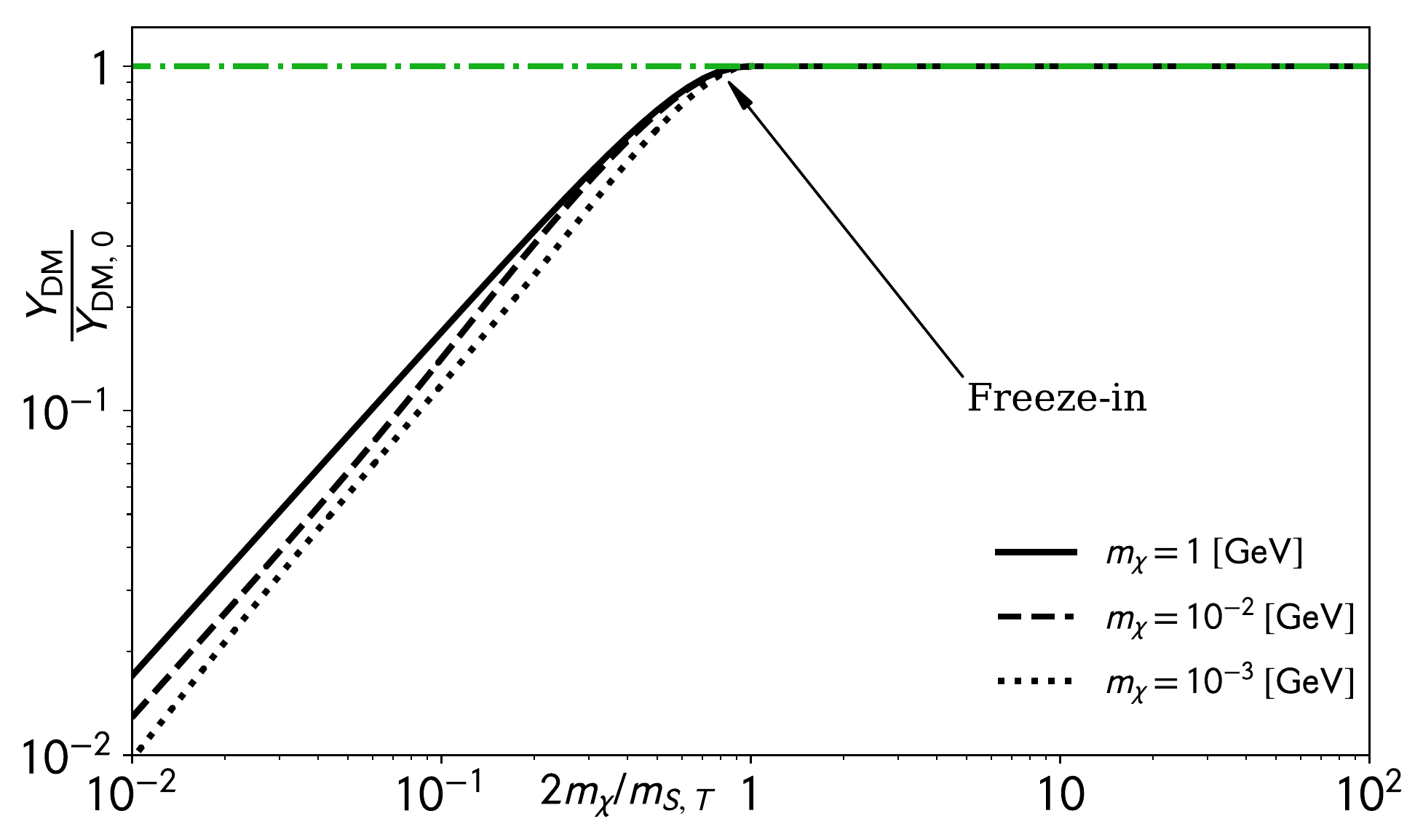}
		\includegraphics[width=0.5\textwidth]{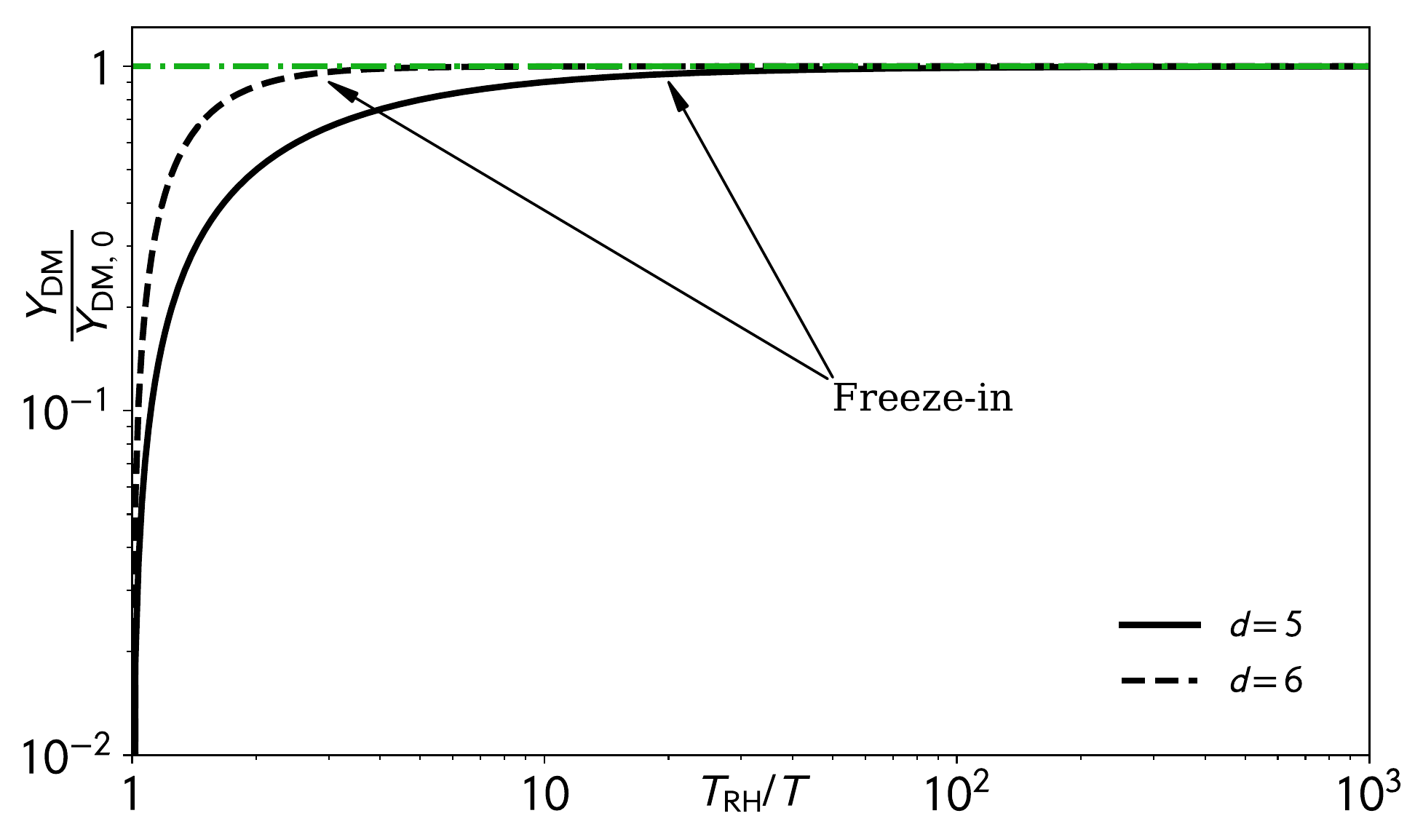}}
	\caption{A typical evolution of the DM yield for the forbidden freeze-in via renormalizable (left) and non-renormalizable (right) operators. }
	\label{Fig:Y_FFI}
\end{figure}
The two solutions of~\eqs{eq:BE_FFI} are shown in Fig.~\ref{Fig:Y_FFI}. The left figure shows the evolution of the yield assuming the renormalizable interactions while the
right one shows the production via non-renormalizable operators.

\section{Standard vs forbidden freeze-in} 
To complete the analysis, we append some results that show the difference between the standard and forbidden freeze-in.
We consider again  the Lagrangian terms of~\eqs{eq:LSchi} without assuming  dominance of either the vacuum or the thermal contribution to the mass of the mediator. 
This will allow us to examine the entire parameter space in both cases, as well as take into account the  case where there is no clear dominance.

\begin{figure}[htb]
	\centerline{%
		\includegraphics[width=0.5\textwidth]{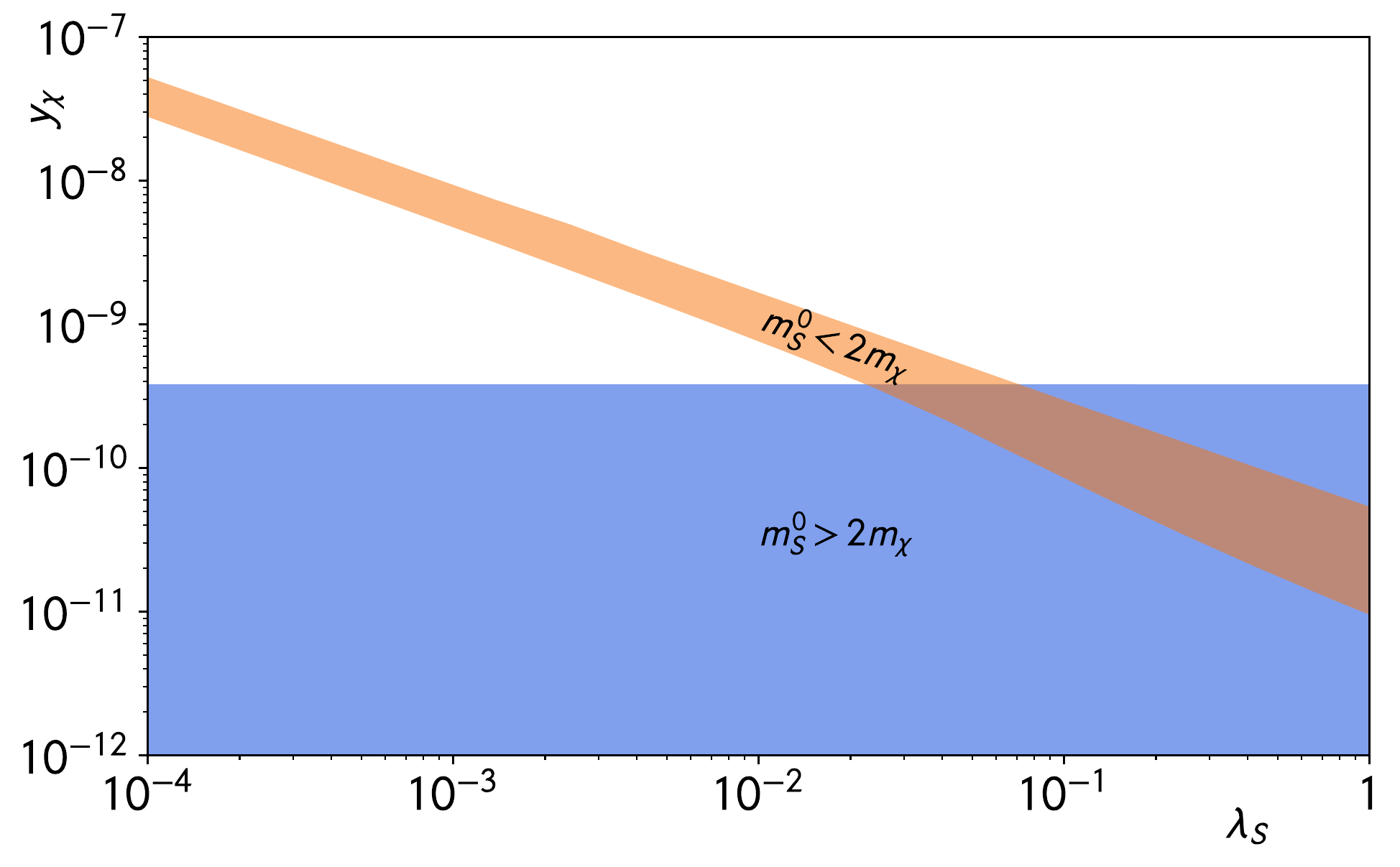}
		\includegraphics[width=0.5\textwidth]{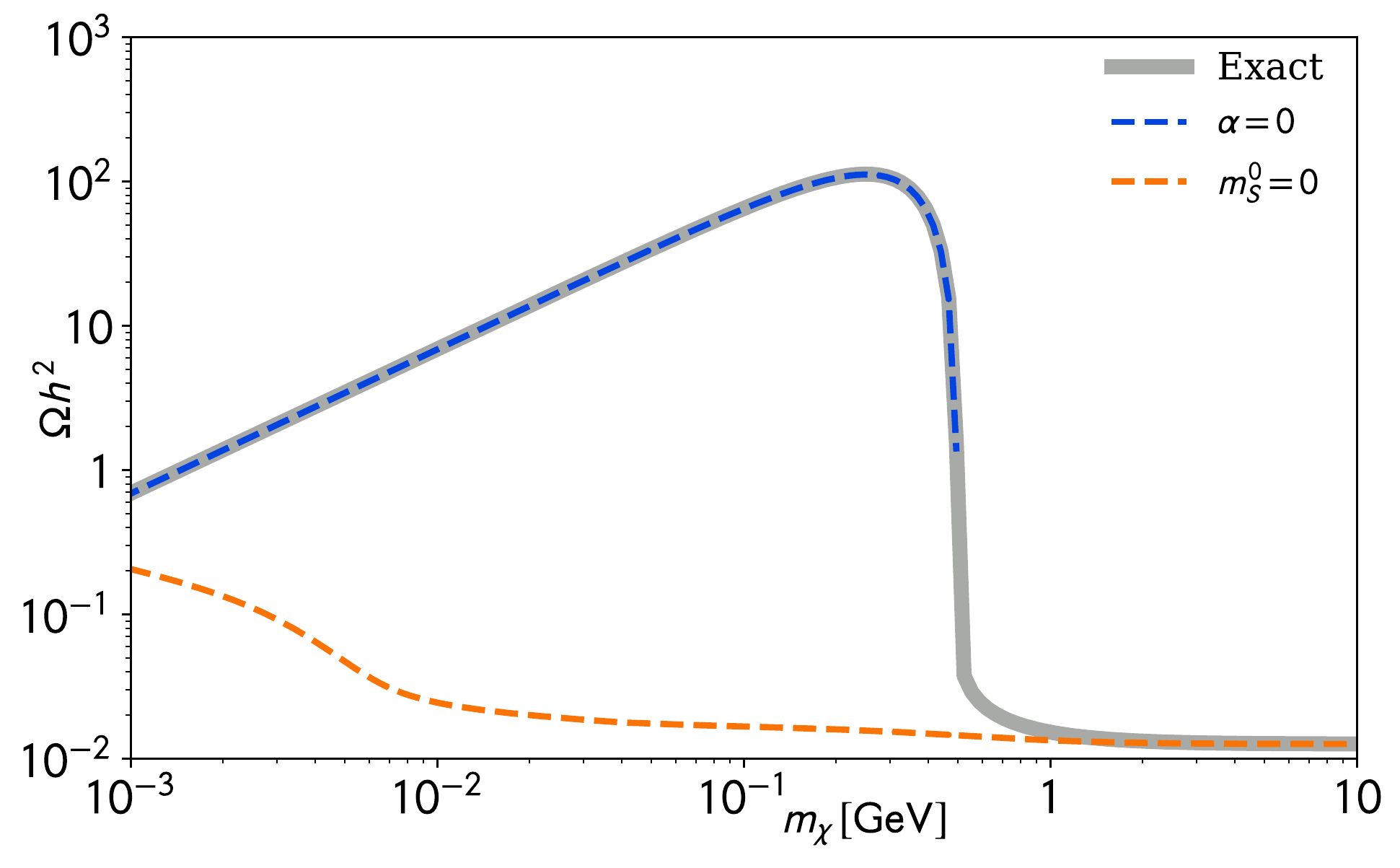}}
	\caption{The parameter space in the $\lambda_S - \ysx$ plane (left) and the dependence of $\Omega h^2$ on $\mc$ (right) given by the numerical solution of the BE~(\ref{eq:BE_std}), assuming thermal mass for $S$. }
	\label{Fig:std_vs_FFI}
\end{figure}
Taking the mediator mass as in~\eqs{eq:mST}, we can solve~\eqs{eq:BE_std} numerically. In the left plane of Fig.~\ref{Fig:std_vs_FFI} we show parameter space in the 
$\lambda_S - \ysx$ plane (left)  that produces the observed relic abundance~\cite{Aghanim:2018eyx}. .
The region of the parameter space the corresponding to the forbidden freeze-in is delineated in orange, while the blue region corresponds to the standard freeze-in case.  
As expected, the parameter space region corresponding to the standard freeze-in is independent of $\lambda_{S}$, while the forbidden regime produces (for the most part) a 
distinct region in the parameter space. Especially for small self-interaction coupling, the forbidden production becomes inefficient (as indicated by~\eqs{eq:YIR_FFI}), and a 
larger Yukawa coupling is needed. 
In right plane of Fig.~\ref{Fig:std_vs_FFI} the dependence of the relic abundance on $\mc$ is shown. The blue line corresponds the standard freeze-in case (ignoring thermal 
mass corrections), the orange line corresponds to the purely forbidden case (with $\mSV = 0$), while the numerical solution (using the mass of~\eqs{eq:mST}) is shown in gray.
In this figure, it is clearly shown that, for the same couplings, the two cases give completely different results, with the forbidden freeze-in producing much lower relic abundance. 
Furthermore, we observe  that at high enough $\mc$ the relic abundance becomes  independent of $\mc$, which is something that does not happen in the standard case 
(at least for DM production via decays).  

\section{A portal model}
\begin{figure}[htb]
	\centerline{%
		\includegraphics[width=0.5\textwidth]{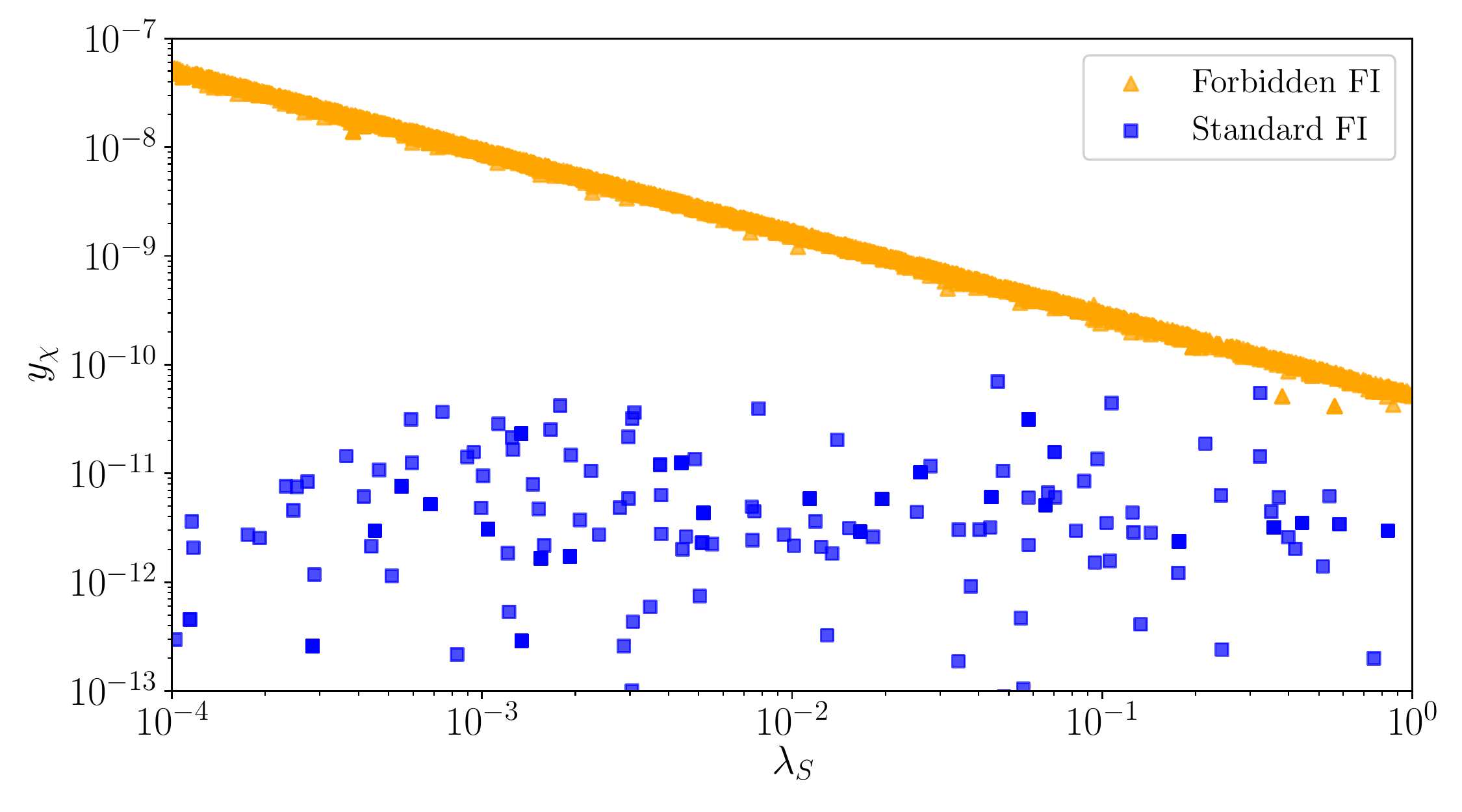}
		\includegraphics[width=0.5\textwidth]{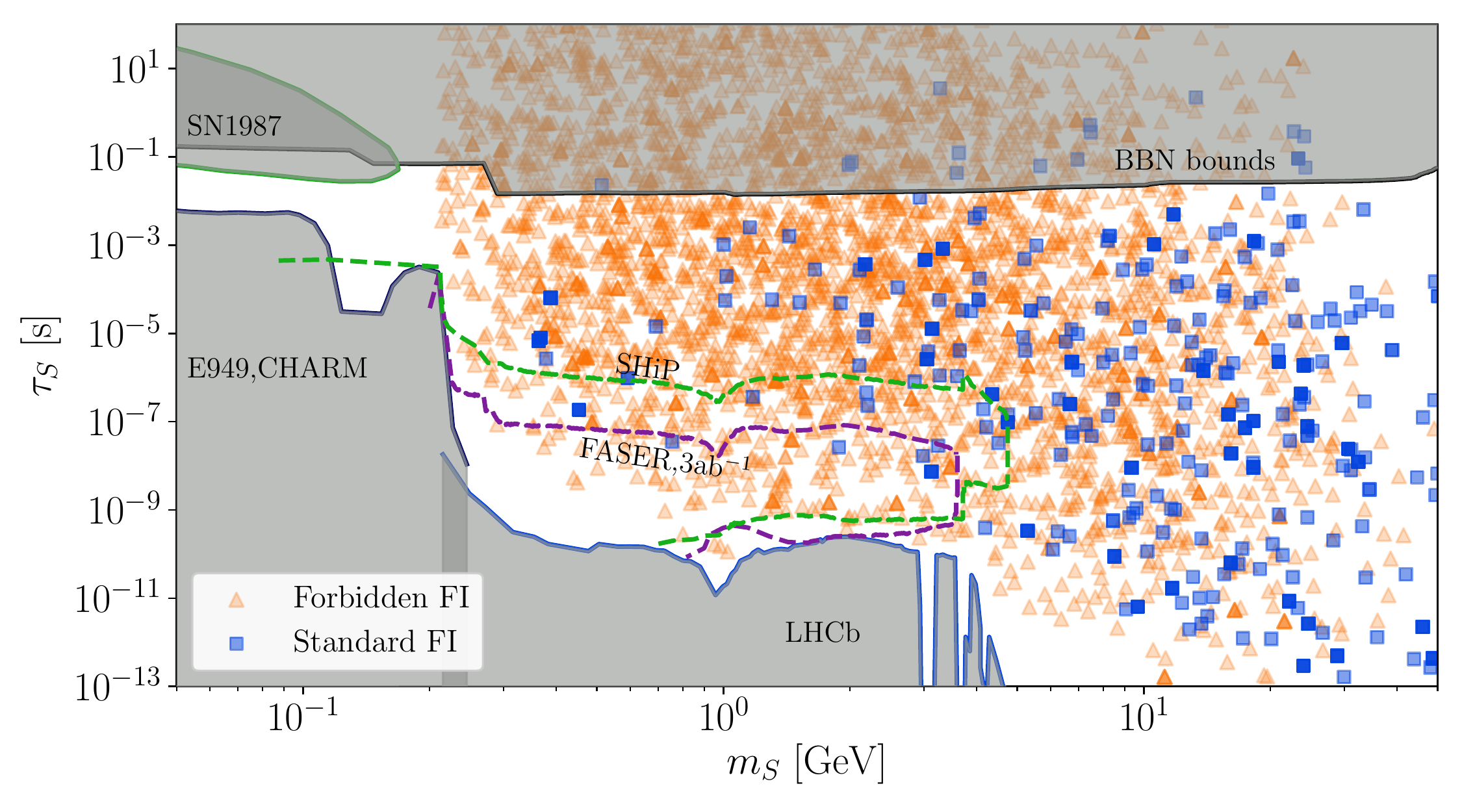}}
	\caption{The parameter space in the $\lambda_S - \ysx$ plane (left) and the (vacuum) mass vs life-time of $S$ (right).  }
	\label{Fig:Portal}
\end{figure}

In the previous section we have treated the dark sector as generally as possible. In a  realistic scenario, though, the $S$  interaction with the Higgs have to be included to correctly 
take into account the decoupling of $S$. 
This is needed especially in the case where $S$ can be decoupled at high energies (which can happen if $S$ is much lighter than the Higgs), which may lead to a very different allowed parameter space than the one shown in Fig.~\ref{Fig:std_vs_FFI}. In this section we show the allowed parameter space (in the $\lambda_S-\ysx$ plane) forbidden freeze-in works in a  ``portal"  model
\begin{align*}
\mathcal{L}^{\rm DM} & =\bar{\chi}\lrb{ i \gamma_\mu \partial^\mu-m_{\chi}  } \chi +  \frac{1}{2} (\partial^\mu S ) (\partial_\mu S)  
-\ysx    S \bar{\chi}  \chi  -  V_{HS}       \ , 
\end{align*}
with the potential
\begin{equation*}
V_{HS}= \dfrac{\mu_{S}^{2}}{2} \, S^2 +  \dfrac{\lambda_{S}}{4!} \, S^4 + 
A \, S \, H^{\dagger} H + \lambda_{HS} \, S^2 \, H^{\dagger} H \, .
\end{equation*}
Solving the system of BEs describing the Higgs, $S$, and DM, we can find the parameter space that gives us the correct relic abundance shown in Fig.~\ref{Fig:Portal}.
We observe that the more realistic case produces a similar allowed parameter space to what is expected according to the discussion of the previous section.  
The inclusion of the Higgs-$S$ interaction, however, introduces a mixing between these particles, which generates couplings with the SM, resulting to a part of the  
parameter space  (mainly the forbidden regime) that violates bounds from big bang nucleosynthesis~\cite{Fradette:2017sdd}. Furthermore there seems to be a region in the 
parameter space that may be probed by future experiments SHiP~\cite{Alekhin:2015byh} and FASER~\cite{Feng:2017vli}.

\section{Conclusions}
Summing up, we have considered production of DM via kinematically forbidden decay of a mediator particle channel, the forbidden freeze-in scenario. 
We have presented a general treatment of this scenario assuming both  renormalizable and  non-renormalizable operators responsible for the decay of the  mediator particle.
We have shown the forbidden regime produces a (mostly) distinct parameter space. 

Also, we have considered a Higgs portal model, which shows that the forbidden freeze-in can be applied to a realistic scenario, which turned out to be also testable in future experiments.

Closing, we point out that this production regime appears to be generic in every frozen-in DM model, since the DM particles are produced from particles in thermal equilibrium with the plasma, which develop thermal masses that increase with the temperature. This results to forbidden decays that open-up at high enough temperatures. Since this type of production is largely neglected in the literature, more models need to be re-examined, in order to identify their forbidden regime.

\bibliography{refs}{}
\bibliographystyle{abbrv}

\end{document}